\documentclass[twocolumn,prl,superscriptaddress,longbibliography]{revtex4-1}

\usepackage{amsmath}
\usepackage{graphicx}
\usepackage{amssymb}
\usepackage{times}
\usepackage{color}
\usepackage{multirow}
\definecolor{lr}{rgb}{1.0,0.3,0.3}
\definecolor{dg}{rgb}{0.0,0.5,0.0}

\usepackage{tabularx} 

\usepackage{comment}

\graphicspath{./}
 


\begin{document}

\title{Isotope purification induced reduction of spin relaxation and spin coherence times in semiconductors}

\author{Oscar Bulancea-Lindvall} 
\affiliation{Department of Physics, Chemistry and Biology, Link\"oping
  University, SE-581 83 Link\"oping, Sweden}
  
\author{Matthew Travis Eiles} 
\affiliation{Max-Planck-Institut f\"{u}r Physik komplexer Systeme, N\"{o}thnitzer Street 38, D-01187 Dresden, Germany}

\author{Nguyen Tien Son} 
\affiliation{Department of Physics, Chemistry and Biology, Link\"oping
  University, SE-581 83 Link\"oping, Sweden}

\author{Igor A. Abrikosov} 
\affiliation{Department of Physics, Chemistry and Biology, Link\"oping
  University, SE-581 83 Link\"oping, Sweden}

\author{Viktor Iv\'ady}
\email{viktor.ivady@liu.se}
\affiliation{Department of Physics, Chemistry and Biology, Link\"oping
  University, SE-581 83 Link\"oping, Sweden}
\affiliation{Max-Planck-Institut f\"{u}r Physik komplexer Systeme, N\"{o}thnitzer Street 38, D-01187 Dresden, Germany}

\date{\today}

\begin{abstract}

Paramagnetic  defects and nuclear spins are often the major sources of decoherence and spin relaxation in solid-state qubits realized by optically addressable point defect spins in semiconductors.   It is commonly accepted that a high degree of depletion of nuclear spins can enhance the coherence time by reducing magnetic noise.   Here we show that the isotope purification beyond a certain optimal level becomes contra-productive, when both electron and nuclear spins are present in the vicinity of the qubits.  Using state-of-the-art numerical tools and considering the silicon vacancy qubit in  various spin environments,  we demonstrate that the coupling to spin-1/2 point defects in the lattice can be significantly enhanced by isotope purification. The enhanced coupling shortens the spin relaxation time that in turn may limit the the coherence time of spin qubits.  Our results can be straightforwardly generalized to triplet point defect qubits, such as the NV center in diamond and the divacancy in SiC.
\end{abstract}
\maketitle


\section{Introduction}

Point defect qubits in  semiconductors exhibit long coherence time  at cryogenic and room temperature.\cite{Balasubramanian:NatMat2009,Christle2014,simin_locking_2017}  Combining this feature  with advanced magneto-optical control of the qubit state has enabled these  systems to become the leading contender in several areas of quantum technology.\cite{Weber10,DegenRMP2017,awschalom_quantum_2018} The properties of point defect qubits depend to a very large degree on the host material.  In particular, magnetic fluctuations in the local spin environment of the defects can profoundly influence the defect's coherence time, in most cases reducing it by several orders of magnitude from the theoretical upper limit set by the spin relaxation time.\cite{Balasubramanian:NatMat2009,Childress:Science2006} As a result, a strategy of chemical and isotope purification in the host material is commonly pursued in order to enhance the coherence time of the point defect qubits.  It is believed that this strategy can be applied in most of the cases.\cite{Childress:Science2006,Balasubramanian:NatMat2009,Mizuochi:PRB2009,bar-gill_solid-state_2013,yamamoto_extending_2013,herbschleb_ultra-long_2019,anderson_five-second_nodate}

In this Letter,  we report on a counterintuitive effect that emerges when both electron  and nuclear spins are present in the vicinity of the qubits.  In particular,  we show that isotope purification leads to a significant reduction of the spin relaxation time,  due to enhanced cross relaxation effects with other paramagnetic defects of similar fine structure.  This effect in turn sets a reduced upper limit for the coherence time.  We study the phenomenon numerically for the quartet silicon vacancy center in SiC, where the consequences may be of high importance.  Our results can be easily extended to other spin defects, such as the NV center in diamond and the divacancy in SiC. 

The negatively charged silicon vacancy in SiC exhibits a quartet ground state spin with long coherence time\cite{Widmann2014,simin_locking_2017}.  This high spin state has been utilized in quantum sensing applications\cite{kraus_magnetic_2014,Lee2015, Simin2016, Niethammer2016,Anisimov2016,soltamov_excitation_2019,hoang_thermometric_2021,abraham_nanotesla_2021} and to implement a room temperature maser \cite{Kraus2014}.  Furthermore,  the defect's favourable optical properties\cite{udvarhelyi_vibronic_2020,udvarhelyi_spectrally_2020}  and advanced fabrication capabilities\cite{widmann_electrical_2019,lukin_4h-silicon-carbide--insulator_2020,babin_fabrication_2022} make it potentially interesting for novel near-infra-red quantum information processing applications.\cite{Baranov2005,Riedel2012, Kraus2014,nagy_high-fidelity_2019,morioka_spin-controlled_2020,lukin_4h-silicon-carbide--insulator_2020,wang_robust_2021,babin_fabrication_2022} In 4H-SiC, the  V1 and V2 photoluminescence lines and the Tv1-Tv2 electron spin resonance (ESR) signals\cite{Wimbauer97,Mizuochi2002,Orlinski2003,Mizuochi2005,son_ligand_2019} are related to the negatively charged silicon vacancy.  The V1 and V2 center were assigned to the $h$ and $k$ silicon vacancy configurations,  respectively,  by comparing with first principles results\cite{IvadyVSi-4H,davidsson_identification_2019}.   In our numerical studies we consider the V2 center,  which is the most often studied configuration.

There are two main ingredients of the environmental spin bath in SiC.  Natural samples include 4.7\% $^{29}$Si and 1.1\% $^{13}$C spin-1/2 nuclear spins. In addition, the structure of the host material incorporates various paramagnetic defects and impurities, whose  concentrations may vary over several orders of magnitudes depending on the growth conditions,  after growth sample preparation,  and nano-scale fabrication.  Here,  we consider the most common intrinsic spin-1/2 defects,  such as carbon vacancy and carbon antisite-vacancy pair whose concentration can reach 10$^{15}$ cm$^{-3}$ in HPSI 4H-SiC.\cite{nagy_high-fidelity_2019,Son2007} We note that this value may increase by 2-3 orders of magnitude due to irradiation and implantation that are frequently used techniques to create silicon vacancy qubits.


In our study,  we divide the complex ground state spin Hamiltonian of the  quartet silicon vacancy-environmental spin bath system into two parts,
\begin{equation} \label{eq:tot}
H = H_{1} + H_2\text{,}
\end{equation}
where $H_{1}$ and $H_2$ describe one and two-spin interaction terms, respectively.  $H_{1}$ includes the zero-field splitting (ZFS) interaction of the quartet silicon vacancy and Zeeman terms of all the spins in the system, 
\begin{equation}\label{eq:H1}
\begin{gathered}
H_{\text{1}} = D \left( S_{0,z}^2 - \frac{5}{4} \right) +  g_{3\parallel}  \mu_B \frac{S_{0,+}^3 - S_{0,-}^3 }{4i}  B_z +\\  g_{e} \mu_B \sum_{j=0}^{M} S_{j,z} B_z   +  \mu_{\text{N}} \sum_{k=1}^N g_{\text{N},k} I_{k,z} B_z   \text{,}
\end{gathered}
\end{equation}
where $S_{j,z}$ is the $z$ component of the electron spin operator vector $\mathbf{S}_j$ of defect $j$,  $I_{k,z}$ is the $z$ component of the nuclear spin operator vector $\mathbf{I}_k$ of nucleus $k$,  $g_e$ is the g-factor of the electron, $\mu_B$ is the Bohr magneton, $g_{\text{N},k}$ is the nuclear g-factor of either $^{13}$C or $^{29}$Si,  and $\mu_{\text{N}}$ is the nuclear magneton.  Terms with $j=0$ index label the silicon vacancy spin, while $j>0$ indices label the doublet paramagnetic defects in the environment. The ZFS parameter $D$ is equal to $ 35.0$~MHz for the V2 silicon vacancy configuration.\citep{IvadyVSi-4H}  The second term on the r.h.s.\ of Eq.~(\ref{eq:H1}) accounts for a non-vanishing higher order term of the Zeeman interaction of the quartet spin states in C$_{3v}$ symmetry,  where $g_{3\parallel} = 0.6$.\cite{Simin2016,IvadyNPJCM2018} The second term in the Hamiltonian, $H_2$ in Eq.~(\ref{eq:H1}),  accounts for the hyperfine and the dipolar coupling of the spins,
\begin{equation} \label{eq:H2}
\begin{gathered}
 H_{\text{2}} =   \sum_{j=0}^M \sum_{k=1}^{N}  \mathbf{S}_j A_{jk} \mathbf{I}_k  +  \\ \sum_{i=0}^{M}   \sum_{j>i}^{M} \frac{\mu_0}{4 \pi} \frac{g_{e}^2 \mu_{B}^2}{r_{ij}^3} \left( \mathbf{S}_i \mathbf{S}_j - 3 \left(\mathbf{S}_i \mathbf{r}_{ij} \right)\left(\mathbf{S}_j \mathbf{r}_{ij}\right)\right)  + \\   \sum_{k=1}^{N}    \sum_{l>k}^{N} \frac{\mu_0}{4 \pi} \frac{g_{\text{N},k} g_{\text{N},l} \mu_{\text{N}}^2}{r_{kl}^3} \left( \mathbf{I}_k \mathbf{I}_l - 3 \left(\mathbf{I}_k \mathbf{r}_{kl}\right)\left(\mathbf{I}_l \mathbf{r}_{kl}\right)\right)     \text{,}
\end{gathered}
\end{equation}
where $A_{jk}$ is  the hyperfine tensor,  $\mathbf{r}_{ij}$ is the position vector pointing from spin $i$ to spin $j$, and $r_{ij} = \left| \mathbf{r}_{ij} \right|$.   $A_{jk}$ for $j>0$ are mostly unknown as they depend on the paramagnetic defects found in the vicinity of the silicon vacancy.  For simplicity,  we consider only the hyperfine interaction of the silicon vacancy, i.e.\ $A_{jk} = 0$ for $j>0$.  This approximation does not affect our qualitative results, however, our quantitative results are the most accurate for environmental paramagnetic defects of small hyperfine interactions, such the common carbon vacancy in SiC.  The hyperfine coupling tensors $A_{0k}$ are obtained from first principles density functional theory calculations in Ref.~\cite{IvadyPRb2020} and used here as well.  Beyond 15~\AA\ distance from the silicon vacancy,  the Fermi contact term is neglected and only the dipolar hyperfine term is considered.  Local inhomogeneities at the defect sites due to the hyperfine interaction are included in calculations as an effective magnetic field.  Accordingly, the first term on the r.h.s.\  of Eq.~(\ref{eq:H2}) is approximated as $H_{\text{inhomo}} = \Delta S_{0,z}$,  where $\Delta  = \sum_k A_{0k,z} \left\langle I_{k,z} \right\rangle$ is the inhomogeneous splitting, or the nuclear Overhauser field in other contexts.  Here,  the  angular bracket represents  expectation value,  while $A_z = \sqrt{A_{xz}^2 + A_{yz}^2 + A_{zz}^2}$.


In order to numerically study the Hahn-echo coherence time (T$_2$) of the quartet silicon vacancy spin in natural and isotope purified SiC, we employ the second order generalized cluster correlation expansion (gCCE2) method\cite{onizhuk_probing_2021}.  The converged models include $M \approx 1000$ nuclear spins within a sphere of radius $r_{\text{Bath}} $  around the qubit.  For natural nuclear spin abundance models $r_{\text{Bath}}= 50$~\AA\ is used, while for lower abundances $r_{\text{Bath}}$ is increased to keep the average number of environmental spins $M$ fixed.  Nuclear spin pairs are considered within the cut-off radius $r_{\text{dip}} = 6.0$~\AA .  The ensemble coherence function is obtained by averaging over 500 randomly generated spin bath configurations and fitted with an $A \text{exp} \!\left( - \left( t / T_2 \right)^n  \right) $ function to obtain the Hahn-echo coherence time $T_2$.   To study the coherence properties of the silicon vacancy, the $ \left| +3/2 \right\rangle $ and the $ \left| +1/2 \right\rangle $ states are used to implement a qubit.

In order to quantify the dipolar spin relaxation time (T$_1$) of the quartet silicon vacancy spin states in a bath of spin-1/2 electron spins, we utilize the method recently developed in Ref.~\cite{IvadyPRb2020} and briefly reviewed in Ref.~\cite{bulancea-lindvall_dipolar_2021}. The spin bath models consist of $N = 32$ electron spins with varying concentration.  We use the first order cluster approximation, which is suitable for an electron spin bath of short coherence time\cite{IvadyPRb2020,bulancea-lindvall_dipolar_2021}.  The time step of the propagation is set to 1~ps,  while the simulation time is optimized for the considered concentrations and vary between 0.05 and 1~ms.  The coupling strength between the qubit and the spin bath spins are obtained in the point-spin density approximation,  i.e.\  by using the second term on the r.h.s.\ of  Eq.~(\ref{eq:H2}).


\begin{figure}[h!]
\includegraphics[width=0.75\columnwidth]{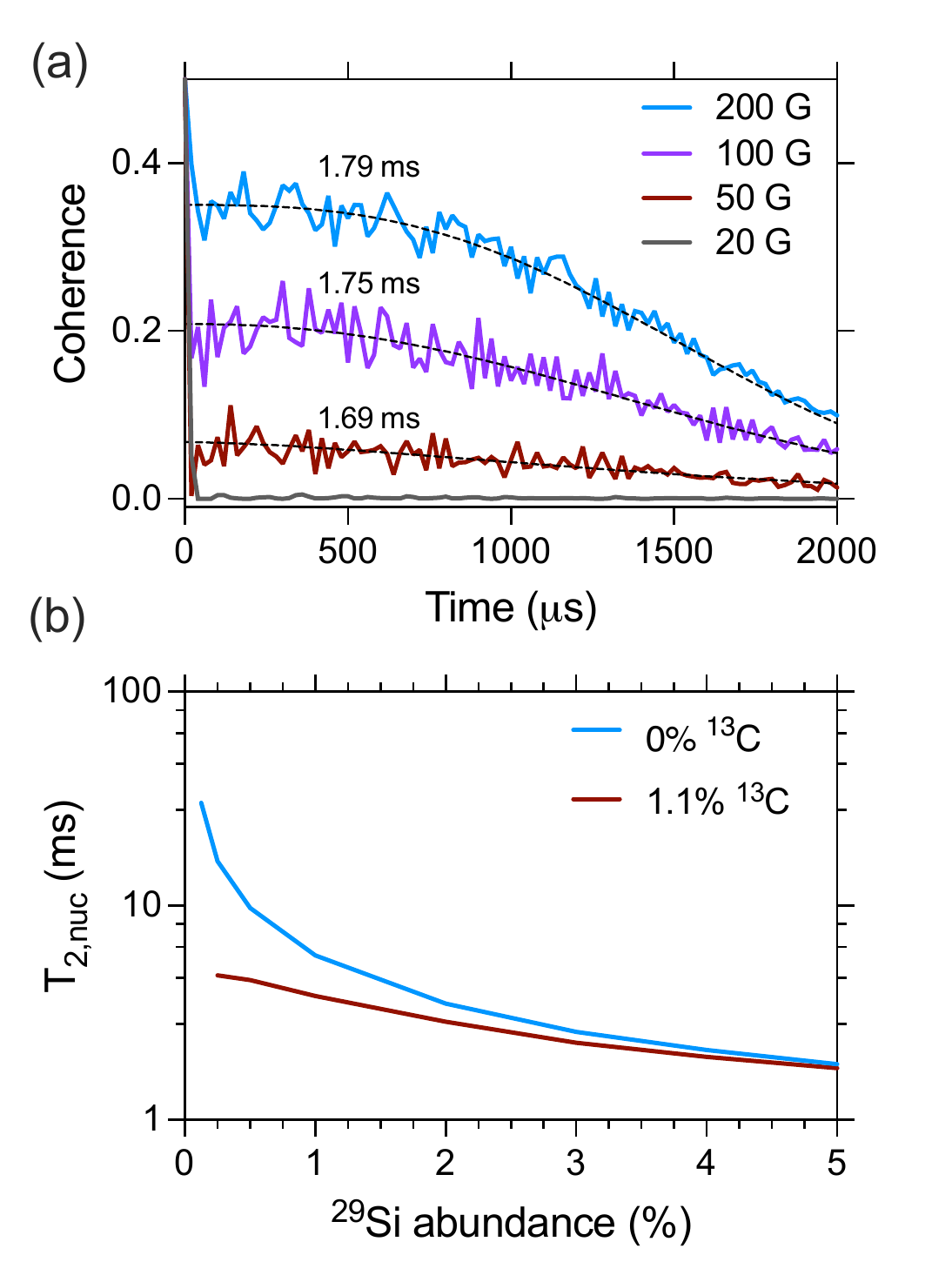}
\caption{  Coherence of the quartet V2 center in a nuclear spin bath in  SiC.  (a) Decay of the coherence function at various magnetic field values for natural abundance of nuclear spins.  The coherence time is provided for each curve.  For $B> 200$~G the coherence time saturates and takes the value of T$_2$ = 1.8~ms.  (b) $^{29}$Si abundance dependence of the saturated coherence time at $B = 200$~G.   
\label{fig:fig1}  }
\end{figure}

First, we study the coherence time of the V2 silicon vacancy qubit when \emph{only nuclear spins} are included in the spin bath.  The decay of the Hahn echo coherence function due to a surrounding nuclear spin bath of natural isotope abundance is depicted in Fig.~\ref{fig:fig1}(a).  As can be seen,  the coherence function decays on two  different time scales.  Due to the hyperfine interaction driven precession of the nuclear spins,  see the first term on the r.h.s.\ of Eq.~(\ref{eq:H2}), the coherence function partially collapses at first with a time scale comparable with the inhomogeneous coherence time ($T_2^{\star}$).  In contrast to the NV center, the coherence function does not recover later and no coherent beatings can be observed.  This irregular behaviour is due to the quartet spin state and further discussed in Ref.~\cite{yang_electron_2014}.  The long time scale decay,  observable in Fig.~\ref{fig:fig1}(a),  is due to the nuclear spin-nuclear spin interaction induced magnetic field fluctuations,  see the last term on the r.h.s.\ of Eq.~(\ref{eq:H2}). The former effect dominates at small magnetic field values, i.e.  at strong hyperfine coupling,  while the latter effect dominates at high magnetic field values where the hyperfine interaction is suppressed by the Zeeman splitting of the nuclear spin states.   The coherence time saturates above $B = 200$~Gauss and takes the values of T$_2$ = 1.8~ms for natural abundance of paramagnetic nuclei.  The strength of the nuclear spin coupling and thus the saturated high magnetic field coherence time sensitively depend on the abundance of the nuclear spins.  As expected,  the T$_2$ time significantly enhances as the nuclear spin bath is depleted,  see Fig.~\ref{fig:fig1}(b).  These results are in accordance with the anticipated behaviour of the system and  previous results on the coherence time of the silicon vacancy.\cite{yang_electron_2014}

\begin{figure}[h!]
\includegraphics[width=0.75\columnwidth]{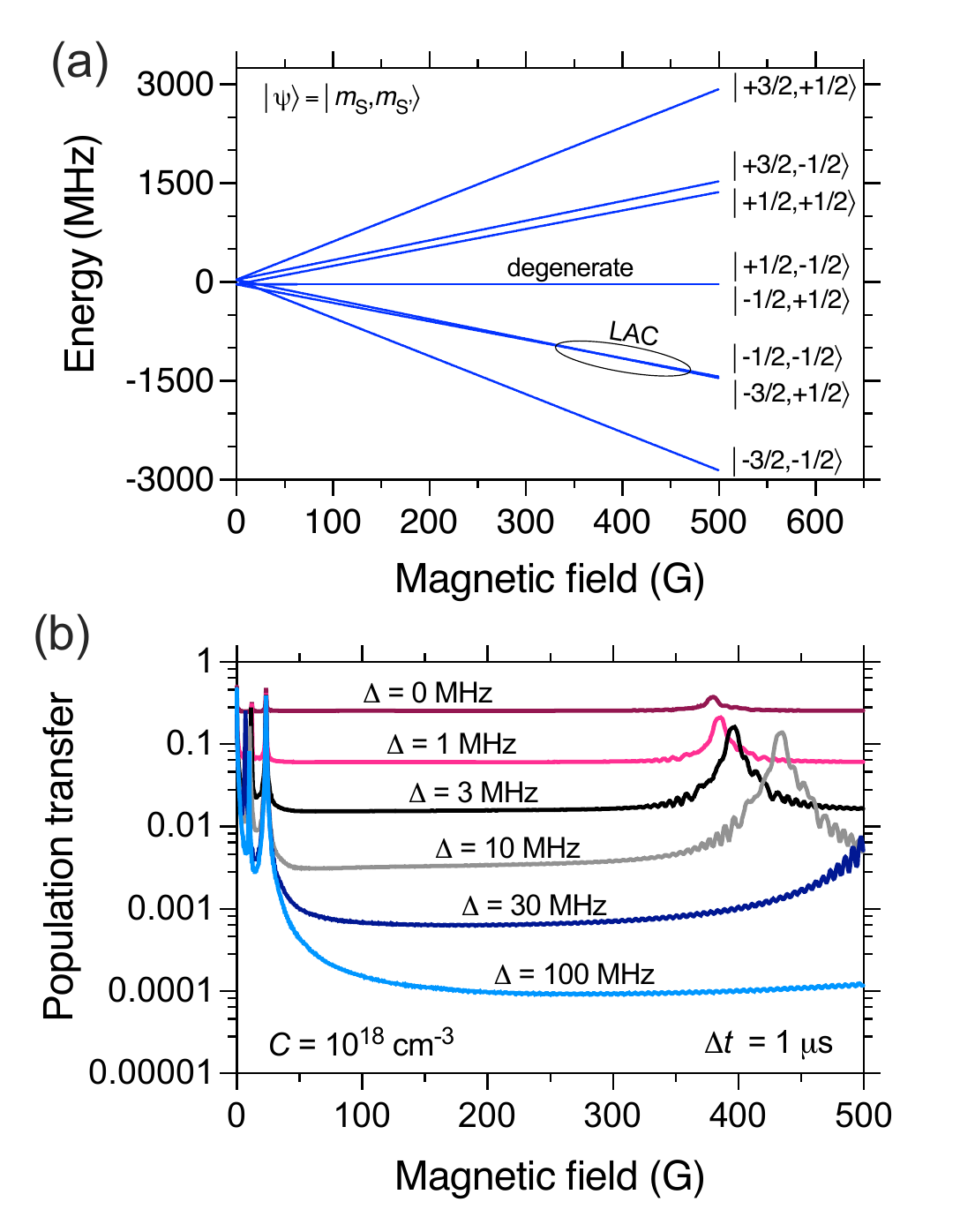}
\caption{  Resonant coupling of a quartet electron spin and a  doublet electron spin.  (a) Energy levels of a quartet-doublet two electron spin system.  The corresponding states are provided in the $\left| m_{\text{S}}, m_{\text{S}^\prime} \right\rangle $ basis, where the quantization axes is set parallel to the $c$-axis. (b) Thermalization of the quartet spin states in a bath of spin-1/2 electron spins.  The vertical axis measures the amount of population transferred from the initial highly polarized $m_{\text{S}}=+1/2$ state to the rest of the quartet states  of the V2 center under a fixed evolution time of $\Delta t = 1$~$\mu$s in a bath of spin-1/2 electron spin of $C = 10^{18}$~cm$^{-3}$ concentration.  The figure depicts polarization transfer curves for various inhomogeneous splitting $\Delta$ of the silicon vacancy states.
\label{fig:fig2}  }
\end{figure}

However,  the host material includes not only nuclear spins but also other \emph{electron spin defects} in the local environment of the qubits.  In order to understand the behaviour of a quartet electron spin in a bath of spin-1/2 defects,  we depict the magnetic field dependence of the energy levels of a quartet-doublet two electron spin system in Fig.~\ref{fig:fig2}(a).  For large magnetic field values the Zeeman interaction is the strongest term in the spin Hamiltonian in Eq.~(\ref{eq:tot}).  Due to the magnetic splitting of both the quartet and the doublet electron spins ($g_e \approx 2$ for both spins), the energy levels form five distinct branches.  The states that are included in the branches are  labelled in Fig.~\ref{fig:fig2}(a). Note that each of the three innermost branches consist of a pair of states. These pairs include $\Delta m_{\text{S}} = \pm 1 $ and  $\Delta m_{\text{S}^{\prime}} = \mp 1 $ states of the quartet and the doublet states  and can be effectively coupled by the $S_{0,+}S^{\prime}_{j,-} + S_{0,-}S^{\prime}_{j,+} $  term of the second term in the r.h.s\ of  Eq.~(\ref{eq:H2}). This interaction induces spin flip-flops of the electron spins, shortens the spin state lifetime,  and thus limits the coherence time of the quartet silicon vacancy qubit states.  

Since the dipolar coupling of the electron spins is generally small,  any splitting of the spin states within the branches has a  significant effect on the lifetime of the states.  In this respect,  it is noteworthy that the $\left | +1/2, -1/2 \right\rangle $ and the  $\left | -1/2, +1/2 \right\rangle $ states are degenerate,  when only ZFS and Zeeman interactions are taken into consideration,  see Fig.~\ref{fig:fig2}(a).  In contrast, the zero-field interaction splits the states in other branches.  Due to the higher order terms of the Zeeman interaction of the quartet silicon vacancy, the states within a branch may cross each other,  see for instance the region labelled by LAC in Fig.~\ref{fig:fig2}(a).

In order to study the thermalization of the silicon vacancy spin states in an electron spin bath,  we simulate the population transfer from the initially highly polarized $\left| +1/2 \right\rangle$ state to the rest of the quartet spin states in Fig.~\ref{fig:fig2}(b).  When the $\left | +1/2, -1/2 \right\rangle $ and  $\left | -1/2, +1/2 \right\rangle $ states are degenerate,  the initial population thermalizes within  1~$\mu$s irrespective of the external magnetic field,  see the uppermost curve in Fig.~\ref{fig:fig2}(b).  On the other hand,  by introducing an effective inhomogeneous magnetic field acting on the quartet spin state and causing a $\Delta$ splitting in the $ m_{\text{S}} = \left\lbrace -1/2, +1/2  \right\rbrace $ subspace of the quartet spin,  the magnetic field independent component of the spin relaxation reduces drastically,  indicating the elongation of the spin state lifetime at most magnetic field values.  The remaining high population transfer peaks, below 50~G and at around 400-500~G in Fig.~\ref{fig:fig2}(b),  are related to level anti-crossings (LACs) of the spin states that are studied in more details in Ref.~\cite{bulancea-lindvall_low-field_2022}.  Furthermore, we note that the reduced lifetime of the $m_{\text{S}} = \left\lbrace -1/2, +1/2\right\rbrace$ state,  beyond the theoretical expectations\cite{soltamov_excitation_2019}, has been recently reported in Ref.~\cite{ramsay_relaxation_2020} and has been attributed to the coupling to spin-1/2 defects.  These results indicate that the inhomogeneous splitting of the spin states plays a crucial role in elongating the spin state lifetime in an electron spin bath.

\begin{figure}[h!]
\includegraphics[width=0.75\columnwidth]{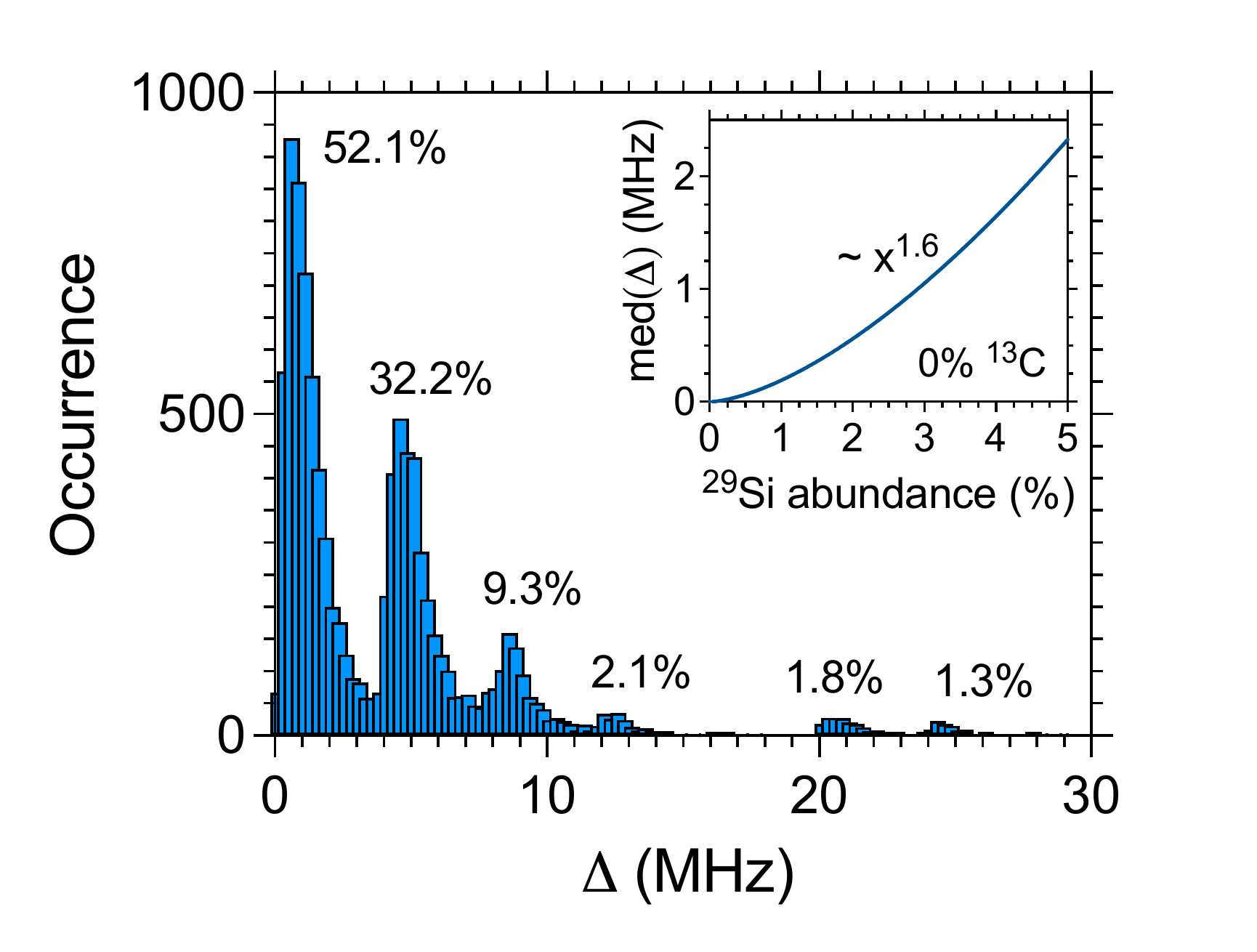}
\caption{  Distribution of the hyperfine splitting  $\Delta$ of the $m_{\text{S}} = \left\lbrace -1/2, +1/2 \right\rbrace$ subspace of the quartet electron spin in natural abundance.   The inset depicts the variation of the median of the hyperfine splitting distribution as a function of the paramagnetic $^{29}$Si abundance.
\label{fig:fig3}  }
\end{figure}

There are several different sources of local inhomogeneous fields that in principle could induce a splitting of the degenerate subspace of the coupled quartet-doublet spin states and  suppress the mutual flip-flops of the spins.  Hyperfine coupling is typically the strongest interaction that can give rise to local inhomogeneities  on sub-nanomater scales.  In order to quantify the hyperfine interaction induced inhomogeneity,  we study the distribution of the hyperfine splitting of the $ m_{\text{S}} = \left\lbrace -1/2, +1/2  \right\rbrace $ subspace of the quartet silicon vacancy.  In Fig.~\ref{fig:fig3}(a), we depict the distribution of the maximal hyperfine splitting $\sum_{k} A_{0k,z}$ obtained in an ensemble of $10^4$ random nuclear spin bath configurations of natural isotope abundance.  There are several peaks in the probability distribution of the maximal hyperfine  splitting.  The largest peak on the left corresponds to configurations with no first and second nearest neighbour nuclear spins.  Going from left to right the second, third, and fourth peaks include configurations with one, two, and three second nearest neighbour $^{29}$Si nuclear spins.  This series continues with vanishing peak heights.  Furthermore,  there are two additional noticeable peaks beyond 20~MHz that correspond to one $^{13}$C and zero $^{29}$Si nuclear spin and one $^{13}$C  and one $^{29}$Si nuclear spins in the first and second neighbourhood shell of the silicon vacancy. 

To study the nuclear spin abundance dependence of the hyperfine interaction induced local inhomogeneity we take the median of the distribution,  for which we obtain 3.97~MHz in natural abundance.  We note that a single value cannot properly characterize a multi peak distribution observed in Fig.~\ref{fig:fig3}(a), however,  in paramagnetic isotope depleted samples the amplitude of the peaks beyond the first peak are significantly reduced and the distribution converges to a single peak-asymmetric distribution.  In such cases, the median is a good measure of the distribution of the maximal hyperfine splitting.  The median as a function of the $^{29}$Si abundance is depicted in Fig.~\ref{fig:fig3}(b). As can be seen,  the median of the hyperfine splitting polynomially approaches zero as the paramagnetic silicon isotopes are depleted. 

\begin{figure}[h!]
\includegraphics[width=0.9\columnwidth]{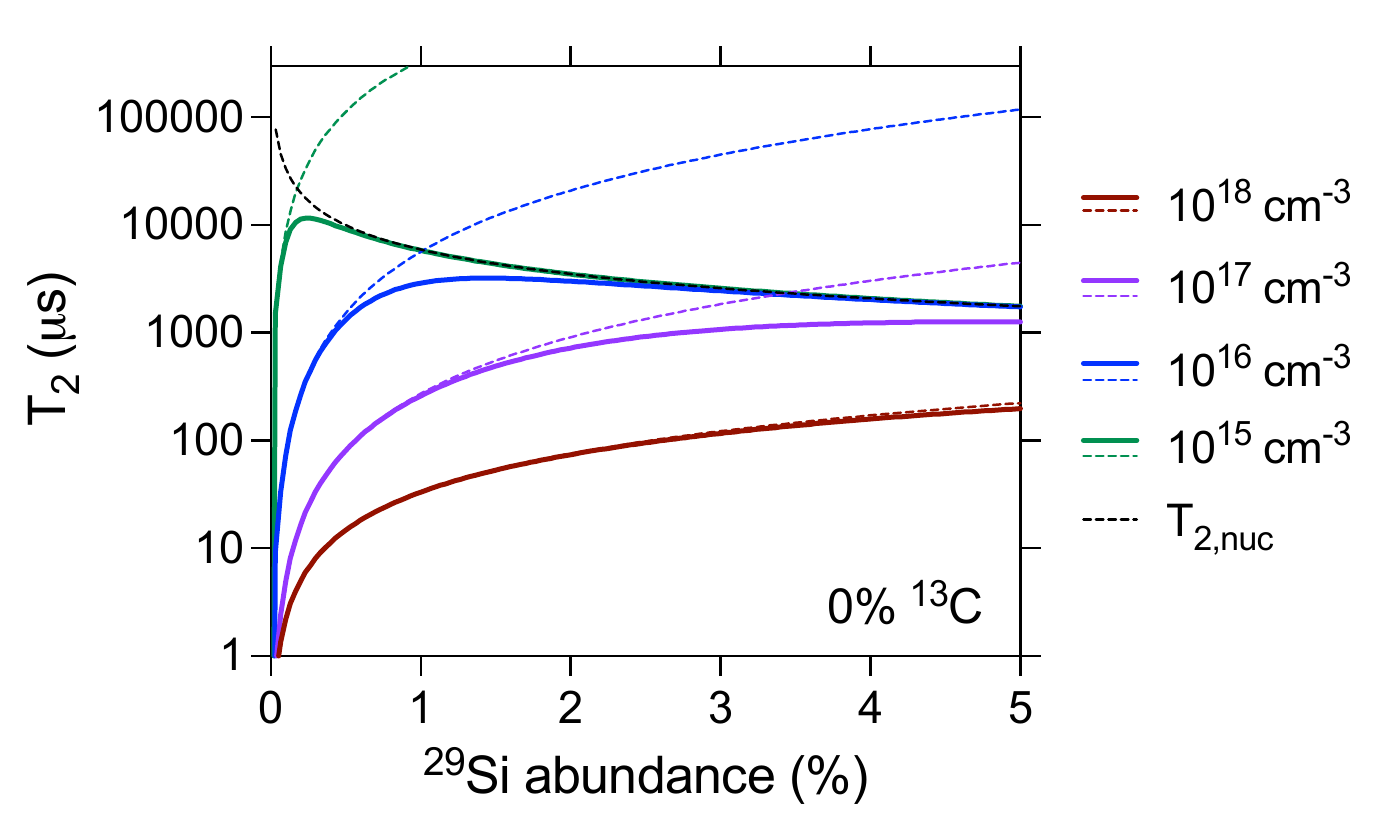}
\caption{  Paramagnetic  $^{29}$Si abundance dependence of the spin coherence time at various spin-1/2 point defect concentrations. T$_{2,\text{nuc}}$ (dashed black line) is obtained at 200~G by including $^{29}$Si nuclear spins in the environment only.   The theoretical maximum of the coherence time T$_{2,\text{max}}$ is set by spin relaxation due to electron spins (plotted by colored dashed lines).  The coherence time T$_2$ is obtained by combining these effects (colored thick solid lines).
\label{fig:fig4}  }
\end{figure}

Combining our results presented so far, we conclude that the isotope purification reduces the inhomogeneous splitting of the qubit states,  see Fig.~\ref{fig:fig3}(b), which in turn can enhance the coupling and  cross-relaxation effects between the quartet silicon vacancy spin states and spin-1/2 defects in the local environment, see Fig.~\ref{fig:fig2}(b).  This counterintuitive phenomena may lead to a drastically reduced spin state lifetime that sets the maximum for the coherence time.  To quantify this effect, we calculate the spin relaxation time T$_1$ of the $\left| +1/2 \right\rangle$ state the silicon vacancy as a function of the $^{29}$Si abundance for various spin-1/2 paramagnetic point defect concentrations.  The hyperfine interaction induced local inhomogeneity, i.e.\ the first term on the r.h.s.\  of Eq.~(\ref{eq:H2}), is approximated by an inhomogeneous magnetic field acting on the silicon vacancy and splits the $m_{\text{S}} = \left\lbrace -1/2, +1/2 \right\rbrace$ subspace by $\Delta = \text{median} \! \left( \sum_k A_{0k,z} \right)$,  see Fig.~\ref{fig:fig3}(b).   We define the absolute maximum of the coherence time as  $\text{T}_{\text{2,max}} = 2 \text{T}_1$.  Here,  we note that in experiments $\text{T}_{\text{2, max}} \approx 0.5 \text{T}_1$ is found for the NV center\cite{bar-gill_solid-state_2013}, therefore our results can be considered as an upper bound.  To obtain the coherence time when both nuclear spins and electron spins are included in the local environment of the silicon vacancy,  we use the $\text{T}_2^{-1} = \text{T}_{2,\text{max}}^{-1} + \text{T}_{2,\text{nuc}}^{-1} $ relation,  where the last term accounts for decoherence effects due to nuclear spin flip-flops induced magnetic fluctuations.  Finally,  we note that direct calculation of the T$_2$ time with the gCCE method provides inaccurate results in this case,  as this method substantially overestimates relaxation effects of strongly coupled spin systems\cite{onizhuk_probing_2021}. 

The results on the ensemble averaged coherence time are depicted in Fig.~\ref{fig:fig4}.  As can be seen the coherence time can be significantly reduced both by the increase of the electron spin concentration and the depletion of the paramagnetic isotopes.  In high spin-1/2 defect concentration ($\approx$10$^{18}$ cm$^{-3}$) the T$_2$ is maximized by the paramagnetic defects and cannot reach higher than $\approx$100~$\mu$s.  As the defect concentration reduces, the theoretical maximum of the coherence time rapidly increases and the fluctuation of the nuclear spin bath starts to limit the coherence time in natural abundance,  see the calculated T$_2$ time at, for instance,  natural abundance of $^{29}$Si isotope in Fig.~\ref{fig:fig4}.  Isotope purification not only reduces magnetic field fluctuations but also enhances cross relaxation effects that may become the major limiting factor in the coherence time in nuclear spin depleted samples, see Fig.~\ref{fig:fig4}.   We note that even a very low concentration of spin-1/2 defect may have a dramatic effect on the coherence time in highly isotope purifies samples.  

Very recently a nanophotonic device integrating V2 qubit with excellent spin properties was realized in Ref.~\cite{babin_fabrication_2022}.  The spin coherence time was found to be 1.39~ms in a high purity isotope purified sample.  The  isotope abundance of the sample was estimated to be $^{28}\text{Si} > 99.85$\% and $^{12}\text{C} > 99.98$\%.   Considering only magnetic field fluctuations due to the residual nuclear spin bath ($\mathtt{\sim}$0.15\% $^{29}$Si), we would expect a coherence time close to 25~ms. The order of magnitude difference indicates that the coherence time is limited by an effect other that nuclear spin flip-flops.  Based on our results,  interaction with electron spins in the lattice is a possible source of decoherence in this experiment.

Our qualitative and quantitative results are obtained for the quartet silicon vacancy in SiC.  However,  the spin Hamiltonian and the results in Fig.~\ref{fig:fig2} can be straightforwardly generalized to spin-1 point defect qubits interacting with other spin-1 environmental defects.  In this case one can also observe nearly degenerate spin states that are coupled by the dipolar interaction.  For spin-1 defects not only the inhomogeneous hyperfine interaction, but also the differences of the ZFS can contribute to the splitting of the coupled states and suppress cross relaxation effects.  Relaxation of the spin states is the most efficient when nearby spin-1 defects of the same kind are coupled to each other, i.e.\ their ZFS parameters are the same.

\section*{Acknowledgments}

We acknowledge support from the Knut and Alice Wallenberg Foundation through WBSQD2 project (Grant No.\ 2018.0071). Support from the Swedish Government Strategic Research Area SeRC and the Swedish Government Strategic Research Area in Materials Science on Functional Materials at Linköping University (Faculty Grant SFO-Mat-LiU No. 2009 00971) is gratefully acknowledged.  N. T. S. acknowledges the support from the Swedish Research Council (Grant No. VR 2016-04068), the EU H2020 project QuanTELCO (Grant No. 862721). The calculations were performed on resources provided by the Swedish National Infrastructure for Computing (SNIC) at the National Supercomputer Centre (NSC).


%

\end{document}